\documentclass[
 ,final     ]{aipproc}

\layoutstyle{6x9}

\usepackage[english]{babel}
\usepackage{epsfig}
\usepackage{hyperref}
\usepackage{amssymb}
\usepackage{amsfonts}
\usepackage{epsf}
\usepackage{rotating}
\usepackage{graphicx}
\usepackage{amsmath}
\usepackage{fancyhdr}
\usepackage{lineno}
\usepackage{subfigure}
\usepackage{babel}
\usepackage{graphics}
\usepackage{geometry}
\usepackage{pstricks}
\usepackage{color}

\let\a=\alpha   \let\b=\beta

\def\a{\alpha}
\def\b{\beta}

\def\ds#1{#1\kern-1ex\hbox{/}}
\def\dsh{h\kern-1.2ex /}

\newcommand{\bea}{\begin{eqnarray}}
\newcommand{\eea}{\end{eqnarray}}

\def\nn{\nonumber}

\def\beq{\begin{equation}}
\def\eeq{\end{equation}}

\def\ba{\begin{eqnarray}}
\def\ea{\end{eqnarray}}

\newcommand{\beqa}{\begin{eqnarray}}
\newcommand{\eeqa}{\end{eqnarray}}

\begin{document}

\title{Massless Scalar Degrees of Freedom in QCD and in the Electroweak Sector from the Trace Anomaly
\footnote{Presented at \emph{QCD@work}, Lecce, June 18-21 2012}
}


\author{Luigi Delle Rose}{}

\author{Mirko Serino}{
  address={ Dipartimento di Matematica e Fisica "Ennio De Giorgi" \\ 
Universit\`{a} del Salento \\ and \\ INFN Lecce, Via Arnesano 73100 Lecce, Italy\\}
}

\begin{abstract}
The interaction of QCD and electroweak sector with gravity is characterized by the generation of a massless pole in a specific form factor present in the 1-loop effective action. We briefly illustrate how to single out this behaviour in perturbation theory, by taking as an example the $TJJ$ vertex, with $T$ denoting the energy momentum tensor of the Standard Model. 
The breaking of scale invariance, due to the trace anomaly, is then related to the appearance of a local degree of freedom in the effective action, which is of dilaton type. 
\end{abstract}

\maketitle


\section{Introduction}

The insertion of the symmetric energy momentum tensor (EMT) on any generic correlator of a given theory defines the coupling of the 
same theory to gravity, describing the response of Green functions involving the theory currents to external gravitational 
perturbations. Mixed correlation functions involving one or more powers of the EMT and gauge currents play, in this respect, a special 
role, since they exhibit the breaking of scale symmetry. 
Such a breaking is induced by renormalization and is related to the appearance of a trace anomaly in these amplitudes.

An analysis of the effective action describing the interaction of a gauge theory with gravity shows that such an action is 
characterized by the appearance of an effective degree of freedom of dilaton type (\cite{Coriano:2012}). 
This is seen as a massless pole in the structure of these Green functions at leading order 
in the gravitational coupling in the $TJJ$ vertex, where $J$ denotes a neutral gauge current.  We intend to summarize these 
findings, which result from a complex but quite direct analysis of these correlators, having been established 
for QED (\cite{Giannotti:2008cv, Armillis:2010cv}), 
QCD \cite{Armillis:2010qk} and for the complete electroweak sector \cite{Coriano:2011zk}.

\section{Perturbative expansion}

We start with few definitions, focusing our discussions only on the case of the graviton/photon/photon ($TAA$) vertex.

We recall that the fundamental action describing the coupling of gravity to the Standard Model is defined by the three contributions 
\beq S = S_G + S_{SM} + S_{I}= -\frac{1}{\kappa^2}\int d^4 x \sqrt{-g}\, R + \int d^4 x
\sqrt{-g}\mathcal{L}_{SM} + \frac{1}{6} \int d^4 x \sqrt{-g}\, R \, \mathcal H^\dag \mathcal H      \, ,
\eeq
where $\kappa^2=16 \pi G_N$, with $G_N$ being the four dimensional Newton's constant and $\mathcal H$ is the Higgs doublet. 
We have denoted with $S_G$ the contribution from gravity (Einstein-Hilbert term) while $S_{SM}$ is the Standard Model (SM) quantum 
action, extended to curved spacetime. $S_I$ denotes the term of improvement for the scalars, which are coupled to the metric via its 
scalar curvature $R$.
The factor $1/6$ should be recognized as giving a conformally coupled Lorentz scalar, the $SU(2)$ Higgs doublet.

We denote with $T_{\mu\nu}$ the complete (quantum) EMT of the electroweak sector of the Standard Model. This includes the contributions 
of all the physical fields and of the Goldstones and ghosts in the broken electroweak phase. Its expression is uniquely given by the 
coupling of the Standard Model Lagrangian to gravity, modulo the terms of improvements, which depend on the choice of the coupling of 
the Higgs doublets. As we have mentioned, we have chosen a conformally coupled Higgs field.  Our computation is performed in the 
$R_\xi$ gauge.
The expression of the EMT is symmetric and conserved. 
It is therefore given by a minimal contribution $T_{Min}^{\mu\nu}$ (without improvement) and the improvement EMT,  
$T_I^{\mu\nu}$, with 
\bea
T^{\mu\nu} = T_{Min}^{\mu\nu} + T_I^{\mu\nu} \,,
\eea
where the minimal tensor is decomposed into
\bea
T_{Min}^{\mu\nu} = T_{f.s.}^{\mu\nu} + T_{ferm.}^{\mu\nu} + T_{Higgs}^{\mu\nu} + T_{Yukawa}^{\mu\nu} 
+ T_{g.fix.}^{\mu\nu} +  T_{ghost}^{\mu\nu}\, .
\eea
The various contributions refer, respectively, to the gauge kinetic terms (field strength, $f.s.$), the fermions, the Higgs, Yukawa, 
gauge fixing contributions ($g.fix.$) and the contributions coming from the ghost sector. 

\section{The $TAA$ case} 

In the $TAA$ case, we introduce the notation $\Gamma^{(AA)\mu\nu\alpha\beta}(p,q)$ to denote the one-loop amputated vertex function 
with a graviton and two on-shell photons. In momentum space we indicate with $k$ the momentum of the incoming graviton and with 
$p$ and $q$ the momenta of the two photons. \\
The amputated correlator is decomposed in the form  
\bea
\Gamma^{(AA)\mu\nu\a\b}(p,q) = \Gamma_{F}^{(AA)\mu\nu\a\b}(p,q) + \Gamma_{B}^{(AA)\mu\nu\a\b}(p,q) + \Gamma_{I}^{(AA)\mu\nu\a\b}(p,q),
\eea
as a sum of a fermion sector (F), a gauge boson sector (B) and a term of improvement denoted as $\Gamma^{\mu\nu\a\b}_{I}$. 
A complete computation gives for the various gauge invariant subsectors the explicit expressions
\bea
\Gamma^{(AA)\mu\nu\alpha\beta}_{F}(p,q) &=&  \, \sum_{i=1}^{3} \Phi_{i\,F} (s,0, 0,m_f^2) \, \phi_i^{\mu\nu\alpha\beta}(p,q)\,, \\
\Gamma^{(AA)\mu\nu\alpha\beta}_{B}(p,q) &=&  \, \sum_{i=1}^{3} \Phi_{i\,B} (s,0, 0,M_W^2) \, \phi_i^{\mu\nu\alpha\beta}(p,q)\,, \\
\Gamma^{(AA)\mu\nu\alpha\beta}_{I}(p,q) &=&  \Phi_{1\,I} (s,0, 0,M_W^2) \, \phi_1^{\mu\nu\alpha\beta}(p,q) + \Phi_{4\,I} (s,0, 0,M_W^2) 
\, \phi_4^{\mu\nu\alpha\beta}(p,q) \,.
\eea
in terms of form factors  $\Phi_{i}$'s (see \cite{Coriano:2011zk} for all of these and the $\phi_i^{\mu\nu\alpha\beta}(p,q)$'s).
The first three arguments of the form factors stand for the three independent kinematical invariants 
$k^2 = (p+q)^2 = s$, $p^2 = q^2 =  0$ while the remaining ones denote the particle masses circulating in the loop. 
We have used the on-shell renormalization scheme. 
In the $TAA$ vertex, the contribution to the trace anomaly in the fermion sector comes from $\Phi_{1\, F}$ which is given by 
\bea
\Phi_{1\, F} (s,\,0,\,0,\,m_f^2) 
&=& 
- i \frac{\kappa}{2}\, \frac{\alpha}{3 \pi \, s} \sum_{f} Q_f^2 \bigg\{- \frac{2}{3} + \frac{4\,m_f^2}{s} 
\nn \\
&& 
- 2\,m_f^2 \,  C_0 (s, 0, 0, m_f^2, m_f^2, m_f^2)\, 
\bigg[1 - \frac{4 m_f^2}{s}\bigg] \bigg\},
\label{taavertex}
\eea
with the sum taken over all the fermions ($f$) of the Standard Model. 
$C_0$ denotes the three-point scalar integral (see \cite{Coriano:2011zk}) and $m_i$ are the fermion masses.
As one can immediately realize, this form factor is characterized by the presence of an anomaly pole
\beq
\Phi^F_{1\, pole}\equiv i \kappa \frac{\alpha}{9 \pi \, s} \sum_{f} Q_f^2
\eeq
which is responsible for the generation of the anomaly in the massless limit. This pattern is typical of any gauge invariant correlator 
of the Standard Model. The other gauge-invariant sector of the $TAA$ vertex, in the electroweak case, is 
the one mediated by the exchange of bosons, Goldstones and ghosts in the loop. We will denote with $M_W$ the mass of the W's.
In this sector the form factor contributing to the trace is
\bea
\Phi_{1\, B} (s,\,0,\,0,\,M_W^2) 
&=& 
- i \frac{\kappa}{2}\, \frac{\alpha}{\pi \, s} \bigg\{ \frac{5}{6} - \frac{2\,M_W^2}{s} 
\nonumber \\ 
&&
+ 2\, M_W^2\, C_0 (s,0,0,M_W^2,M_W^2,M_W^2) \,
\bigg[1 - \frac{2 M_W^2}{s}\bigg] \bigg\}, 
\label{oneb}
\eeqa
which multiplies the tensor structure $\phi_1$, responsible for the generation of the anomalous trace. In this case the anomaly pole is 
easily isolated from (\ref{oneb}) in the form
\beq \label{Phi1Bpole}
\Phi_{1\, B, pole}\equiv - i \frac{\kappa}{2}\, \frac{\alpha}{\pi \, s} \frac{5}{6} \,.
\eeq
The term of improvement is responsible for the generation of two form factors, both of them contributing to the trace. 
They are given by 
\bea
\Phi_{1\, I} (s,\,0,\,0,\,M_W^2) &=& - i \frac{\kappa}{2}\frac{\alpha}{3 \pi \, s} \bigg\{ 1 + 2 M_W^2 \,C_0 (s, 0, 0, M_W^2, M_W^2, 
M_W^2)\bigg\} ,\\
\Phi_{4\, I} (s,\,0,\,0,\,M_W^2) &=&  i \frac{\kappa}{2}\frac{\alpha}{6 \pi }  M_W^2 \,C_0 (s, 0, 0, M_W^2, M_W^2, M_W^2), 
\eea
the first of them being characterized by an anomaly pole 
\beq \label{Phi1Ipole}
\Phi_{1\, I\,\, pole}=  - i \frac{\kappa}{2}\frac{\alpha}{3 \pi \, s}.
\eeq
The same situation is encountered in QCD, when $J$ denotes a gluon current: 
here both the gluon loop and the fermion loop are separately gauge invariant and both exhibit such a structure.

\section{Implications for the effective action}

The pole-like behaviour generated by the anomaly diagrams have a general implication concerning the structure of the effective action. 
We recall \cite{Giannotti:2008cv, Armillis:2010qk} that in the case of massless QED the effective interaction induced by the trace 
anomaly takes the form 
\beq
\mathcal{S}\sim \int d^4 x d^4 y\, R^{(1)}(x) \square^{-1}(x,y) F_{\mu\nu}(y) F^{\mu\nu} (y) 
\label{reig}
\eeq
where $R^{(1)}$ denotes the linearized scalar curvature and $F_{\mu\nu}$ is the abelian field strength. A similar result holds for QCD 
\cite{Armillis:2010qk}.  As shown in \cite{Giannotti:2008cv} this expression coincides with the long-known anomaly-induced action 
obtained by Riegert \cite{Riegert:1984kt}, which was derived for a generic gravitational field, after an expansion of its expression 
around the flat spacetime limit. Notice that in terms of auxiliary degrees of freedom (i.e. two scalar fields $(\varphi, \psi')$) which 
render the action (\ref{reig}) local \cite{Giannotti:2008cv}, extra couplings of the form $\varphi F F$ are automatically induced by 
the $1/\square$ term. This interaction is indeed present in the equivalent Lagrangian 
\beq
S_{anom} [g,A;\varphi,\psi'] =  \int\,d^4x\,\sqrt{-g}
\left[ -\psi'\square\,\varphi - \frac{R}{3}\, \psi'  + \frac{c}{2} F_{\alpha\beta}F^{\alpha\beta} \varphi\right]\,,
\label{effact}
\eeq
($c=-\beta(e)/(2 e)$)
where $\varphi$ and $\psi'$ are auxiliary scalar fields. $\varphi$ describes a dilaton interaction. 

\section{Conclusions}
These findings seem to indicate that the signature of the trace anomaly is in the appearance of a massless degree of freedom of dilaton 
type which force us to draw a parallel between this anomaly and the chiral anomaly. In the chiral case, 
anomaly poles have been a distinctive signature of the anomaly for almost two decades, which in a local formulation of the effective 
Lagrangian amount to axion-like interactions, in the form of Wess-Zumino terms \cite{Armillis:2009im}, studied in the context of 
anomalous $U(1)$ extensions of the Standard Model \cite{Coriano:2005js, Coriano:2007xg}. 
There are several implications of these findings. We have discussed in a related contribution to these proceedings the role of such dilaton-like degree of freedom in the context of conformal extensions of the Standard Model, and its possible appearance, in the broken electroweak phase, together with the Higgs scalar \cite{Coriano:2012}. There are also other implications. We mention, for instance, the case of supersymmetric theories, where the anomaly 
multiplet is known to contain as components the trace of the EMT, an axial anomaly related to the $U(1)_R$ current $(J_R)$ and the gamma-trace of the 
supersymmetric current. Our results clearly imply that such poles are present when we insert two (i.e. $T$ and $J_R$) of these three operators in the effective action. Threfore they should also appear in correlators involving insertions of the supersymmetric current.

\bibliographystyle{aipproc}

\end{document}